\documentclass[twocolumn,prl,amsmath,amssymb]{revtex4}
\usepackage{epsfig}
\begin{document}
\title{Preferential Growth of Pt on Rutile TiO$_2$}
\author{Hakim Iddir, Vladimir Skavysh and Serdar \"{O}\u{g}\"{u}t}
\affiliation{Department of Physics, University of Illinois at Chicago,
Chicago IL 60607}
\author{Nigel D. Browning}
\affiliation{Department of Chemical Engineering and Materials Science,
University of California-Davis, Davis, CA 95616, and NCEM, Lawrence Berkeley
National Laboratory, Berkeley, CA 94720}
\date{\today}
\begin{abstract}

The characterization of Pt/TiO$_2$ (Degussa P25) catalyst system
using atomic resolution Z-contrast images and electron energy loss
spectroscopy in the scanning transmission electron microscope has
recently revealed that Pt particles have a strong tendency to nucleate on
the rutile phase of TiO$_2$ rather than anatase. Comparative 
{\em ab initio} pseudopotential calculations for 
Pt and Pt$_2$ on the stoichiometric and reduced TiO$_2$ surfaces, 
and oxygen vacancy ($V_{\rm O}$) formation energies are performed
to address the microscopic origin of this finding. 
The results, which show that Pt actually binds more strongly to anatase
surfaces, indicate that the selective growth of Pt
on rutile must be controlled by the lower formation energy of $V_{\rm O}$
on rutile, and 
possibly by the stronger tendency of $V_{\rm O}$ sites on rutile
to trap large Pt clusters compared to anatase.

\end{abstract}
\maketitle

The fundamental and technological importance of TiO$_2$, stemming to a large 
extent from its wide spread use as a catalyst and catalyst support,
has made it the subject of many experimental and theoretical 
studies over the last decade.\cite{Diebold} As one of the most 
active catalysts for CO oxidation reactions and photocatalysis, in 
addition to being the prototype 
system for the strong-metal-support-interaction (SMSI) phenomenon,
\cite{Tauster} Pt/TiO$_2$ has recently received particular 
attention. The catalytic properties of Pt/TiO$_2$ and the occurrence
of SMSI has a strong 
dependence on the phase of TiO$_2$ (rutile versus anatase). 
For example, it was recently shown that anatase titania palladium supported
catalyst presents SMSI at low H$_2$ pre-reduction temperatures, while
rutile does not.\cite{Li et al.}
It is also well known that anatase is more efficient
than rutile as an oxidative photocatalyst.\cite{Fu-gra-had-ohno}
The presence of a small 
amount of rutile, however, such as in commercial mixed-phase titania samples 
results in an unusually high activity.\cite{hurum}
A fundamental
comparative study of the interaction of Pt with both rutile and anatase
TiO$_2$ surfaces will,
therefore, contribute to our understanding of the catalytic
properties of this system and the occurrence of SMSI.

\begin{figure}
\begin{center}
\includegraphics[width=0.5\textwidth]{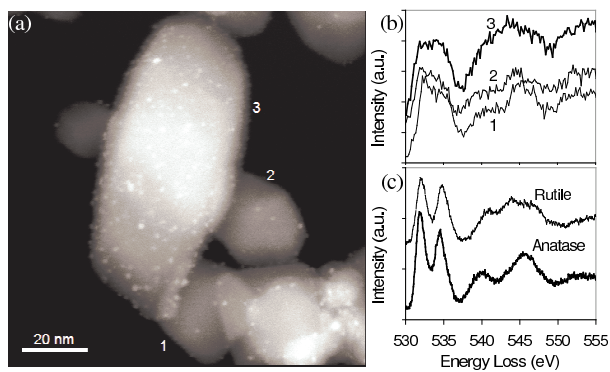}
\end{center}
\vspace{-0.5in}
\caption{(a) Z-contrast image of Pt/TiO$_2$ showing the
preferential growth of Pt (tiny white dots) on rutile. The large
gray particles are TiO$_2$. (b) Oxygen $K-$edge EEL spectra taken 
at the locations specified in (a). Note the difference in the shape
of the secondary peaks (from 538 eV to 548 eV) for particles 1 and 2
in comparison to particle 3. (c) O $K-$edge EEL spectra of bulk rutile and anatase
shown for comparison with the spectra in (b), showing that 
particles 1 and 2 with few Pt clusters are of anatase and 
the densely populated particle 3 is of the rutile phase.}
\end{figure}

Recently, using a combination of Z-contrast imaging and electron
energy loss spectroscopy (EELS) in the scanning transmission electron
microscope (STEM), we examined the atomic and electronic structure
of the Pt/TiO$_2$ interface.\cite{Iddir1} The experiments were performed on 
a commercial mixed-phase titania sample, known as Degussa P25, 
which is composed of about 80 \% anatase and 20 \% rutile. 
We observed rather unexpectedly that Pt particles 
were not uniformly distributed
over the titania particles, but showed a selective distribution, 
as shown in Fig. 1(a). The oxygen$-K$ edges of these specific
particles [Fig. 1(b)] combined with the reference O-$K$ edge
EEL spectra of rutile and anatase TiO$_2$ [Fig. 1(c)] revealed
that the densely Pt populated titania particles were of the rutile
phase while the barely populated ones were of anatase. This 
finding was consistently observed throughout the
samples, and also confirmed with the characteristic peak at around
14 eV in the low loss EELS.\cite{Iddir1,Launay}
Similar observations were reported earlier for
Ru-RuO$_{x}$/TiO$_2$, Ir/TiO$_2$, and Au/TiO$_2$ systems.\cite{Rut-akita}
Motivated by these interesting experimental observations, in this 
paper, we present results from large-scale {\em ab initio} calculations
which provide possible explanations for the observed preferential 
growth of Pt particles on rutile rather than anatase.

Our calculations for the atomic and electronic structures of 
single Pt atoms and Pt$_2$ dimers
on the stoichiometric and reduced rutile and anatase TiO$_2$
surfaces were performed using the pseudopotential 
total energy method\cite{vasp}
in a slab geometry. The specific surfaces chosen for this study,
(110) for rutile and (101) for anatase, are the most energetically
stable surfaces of each phase forming a large portion of surface
area in Wulff construction studies.\cite{Rama-Laz}
We used ultrasoft pseudopotentials with
a cutoff energy of 300 eV, $1\times2\times2$ Monkhorst-Pack
k-point grids, and Perdew and Wang parametrization of the
generalized gradient approximation.
Increasing the cutoff energy 
to 400 eV and the k-point grid to $2\times4\times4$ had no appreciable
effect on the results.
We made an
extensive study of the effects of spin polarization effects and
found them to be negligible for the Pt/TiO$_2$ system. After a
systematic study and comparison with larger slabs (checked up to
10 layers in $2\times 1$ surface cells), we found a 4-layer asymmetric
slab, in which the two bottom layers are kept at bulk positions,
to be the smallest good representation of the surface with and
without Pt atoms on the surfaces. We used a vacuum region of 11-12 \AA.
To model the metal atoms on the rutile and anatase surfaces, 
we used large surface unit cells, $3\times 2$ (stoichiometric) 
and $4\times 2$ (reduced), resulting in 144 and 192 atoms per slab, 
respectively. 
More details about the calculations and convergence tests
on rutile (110) can be found 
in Ref. \onlinecite{Iddir2}.

\begin{figure}
\vspace{-1.3in}
\begin{center}
\includegraphics[width=0.5\textwidth]{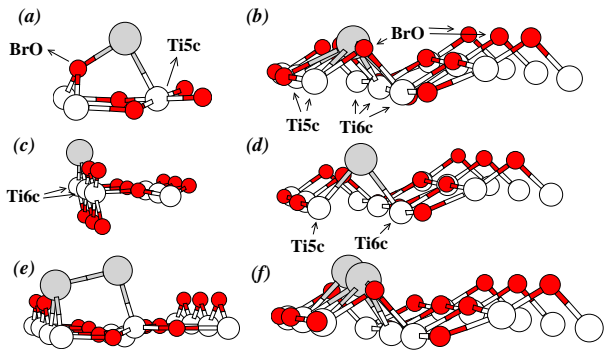}
\end{center}
\vspace{-1.5in}
\caption{The most favorable adsorption sites for Pt on 
(a) stoichiometric rutile (110), (b) stoichiometric anatase (101), 
(c) reduced rutile (110), and (d) reduced anatase (101). Also shown 
in (e) and (f) are the most favorable geometries for Pt$_2$ adsorbed
on reduced rutile (110) and anatase (101), respectively. 
The white, black, and gray  circles
represent Ti, O, and Pt  atoms, respectively. BrO denotes bridging
oxygen.}
\end{figure}

A reasonable explanation for the preferential growth of Pt on rutile
could come from the differences in the binding energies (BEs) of Pt atoms 
to the rutile (110) versus anatase (101) surfaces. That is, based on the 
experimental results one might expect that Pt would bind
more strongly to stoichiometric and/or reduced rutile (110)
compared to anatase (101) surfaces. In order to test this hypothesis,
we first investigated the binding of a single Pt atom to the
stoichiometric rutile (110) and anatase (101) surfaces (Table I).
As reported previously,\cite{Iddir2} we find the most favorable position for
Pt adsorption on the stoichiometric rutile surface as the hollow
site [Fig. 2(a)] with a BE of $E_{S,{\rm Rut}}=2.21$ eV.
At this site Pt binds to a five-fold coordinated Ti atom (Ti5c) 
and a bridging oxygen. 
On the stoichiometric anatase surface, the most favorable site 
for Pt adsorption is the equivalent site to the hollow position
found above [Fig. 2(b)], as Pt binds to the under-coordinated
Ti (Ti5c) and O atoms as well as six-fold coordinated Ti6c.
The calculated BE on the stoichiometric anatase surface
is $E_{S,{\rm Ana}}=2.87$ eV, which is surprisingly higher
by 0.66 eV compared to $E_{S,{\rm Rut}}$. 
By itself, this result would suggest that 
on stoichiometric surfaces, a Pt
atom would prefer to bind to anatase rather than rutile, which
is contrary to our experimental finding of preferential growth
on rutile. 

Since oxygen vacancies are known to play an important role in the 
anchoring of metal atoms on TiO$_2$, we also examined the binding
of Pt atoms to the reduced rutile and anatase surfaces (Table I). As expected,
we find the most favorable adsorption on both 
surfaces to occur at the substitutional (oxygen vacancy) site
[Figs. 2(c) and (d)]. The calculated BEs
of $E_{R,{\rm Rut}} = 3.59$ eV and $E_{R,{\rm Ana}} = 5.05$ eV
are considerably larger compared to those on stoichiometric 
surfaces. However, the surprising result of even stronger preference
of Pt to bind to reduced anatase (101) rather than reduced rutile (110)
by a significantly large BE difference of 1.46 eV
is again contrary to our experimental finding.

At this point, it is natural to ask why Pt binds more strongly
to anatase (101) surfaces than to rutile (110).
We first note that our analysis of the 
charge density profiles to examine the nature of the
bonding of Pt to rutile and anatase surfaces has not
revealed a significant difference between the two phases.\cite{note}
The reasons behind the stronger binding of Pt to anatase
in comparison to rutile can be more easily understood from the different
coordinations of Pt and the resulting effect of these on the Pt $d-$band. 
On the stoichiometric rutile (110), Pt bonds to only
one bridging oxygen (at 2.01 \AA) and one 5-fold 
coordinated Ti atom (at 2.50 \AA), as shown in Fig. 2(a).
On the stoichiometric anatase (101), on the other hand, Pt
fits well (with a very small protrusion above the surface)
into the large spacing between two 2-fold coordinated oxygens,
two 5-fold coordinated Ti, and two 6-fold coordinated Ti atoms [Fig. 2(b)].
In addition to the increased coordination of Pt, the other significant
difference between the two cases is the Pt-Ti interatomic distance, 
which increases to an average of 2.78 \AA\ on anatase (101). 
Examination of the Pt-Ti interatomic distance in various PtTi
compounds, such as in $B2$ and $B19$ structures,\cite{karin} 
shows that the preferred nearest-neighbor Pt-Ti distance 
is near 2.75 \AA\ (the same analysis for Pt oxides shows 
a preferred Pt-O bond near 2.0 \AA). Therefore, on anatase (101)
Pt not only has a higher coordination than on rutile (110), which 
indicates increased binding, it can also position itself at preferred 
distances from both O and Ti due to the larger spacings between the 
bridging oxygen rows. This stronger interaction of Pt with anatase (101) surface
manifests itself in the $d-$partial density of states (PDOS) around
the Pt atom as shown in Figs. 3(a) and 3(b). The stronger hybridization
of the Pt $d-$states with Ti pushes the occupied states (around $-2$ eV in rutile)
down to much lower energies in anatase, while increasing the energy of the unoccupied
states (near 0.6 eV on rutile) to higher energies in anatase, as shown 
by the arrows in Fig. 3(a).

\begin{figure}
\vspace{-0.7in}
\begin{center}
\includegraphics[width=0.5\textwidth]{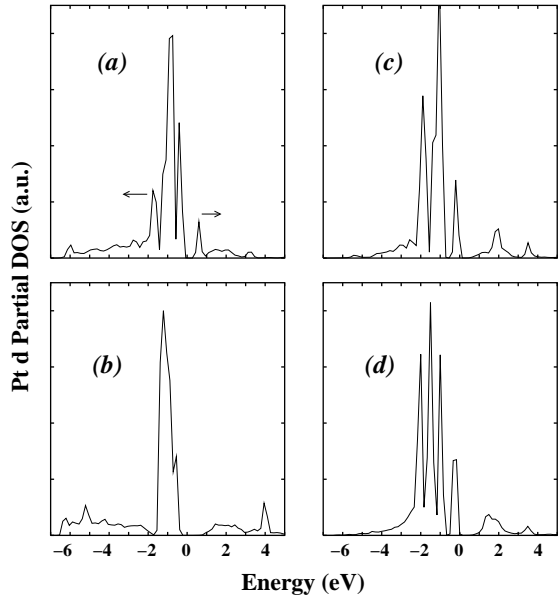}
\end{center}
\vspace{-0.8in}
\caption{Partial $d$ density of states around the Pt atom adsorbed on 
(a) stoichiometric rutile (110), (b) stoichiometric anatase (101),
(c) reduced rutile (110), and (d) reduced anatase (101). The arrows
refer to the discussion in the text.}
\end{figure}

The reason behind the stronger binding of Pt to the reduced anatase (101)
compared to rutile (110) is the difference in the coordination of the two 
Ti atoms bonded to the Pt atom.
On rutile, Pt sits equidistant from
two Ti6c atoms at 2.39 \AA, while on anatase, it bonds to 
two inequivalent Ti atoms, Ti5c and Ti6c, at distances of
2.34 \AA\ and 2.50 \AA, respectively. As a result of this asymmetry
on reduced anatase, there are three main peaks in the Pt $d-$PDOS of 
anatase (101) near the valence band maximum (VBM) compared to two for 
Pt on reduced rutile, as shown in Figs. 3(c) and 3(d). 
The split peaks at $-1.6$ eV and $-1$ eV on anatase
are associated with the interaction of Pt with Ti5c and Ti6c, respectively. 
As a result of the stronger interaction of Pt with Ti5c 
the split $d-$band is pushed down to lower energies and slightly increases 
the width of the Pt $d-$band. 

\begin{table}
\caption{Calculated binding energies per atom (in eV) of Pt and Pt$_2$
on stoichiometric ($S$) and reduced ($R$) rutile (110) and anatase (101)
surfaces.}
\begin{ruledtabular}
\begin{tabular}{lcccc}
& Rutile, $S$ & Anatase, $S$ & Rutile, $R$ & Anatase, $R$\\ \hline
Pt & 2.21 & 2.87 & 3.59 & 5.05 \\
Pt$_2$ & $-$ & $-$ & 3.69 & 4.71 \\
\end{tabular}
\end{ruledtabular}
\end{table}

So far, our
calculations show that (i) Pt binds more strongly to both rutile and
anatase {\em reduced} surfaces than to the stoichiometric ones, and
(ii) Pt binds more strongly to anatase than to rutile on both types of 
surfaces. Since it is clear that the binding energies alone cannot
explain (in a sense, contradict) 
our experimental observation, another factor must play an
important role in the selective growth of Pt on TiO$_2$. The
stronger binding of Pt to the reduced surfaces shows that oxygen
vacancy plays the important role of a strong anchoring site
for adsorption. The next logical step is, therefore, to
investigate how easy it is to create these nucleation sites on
rutile (110) versus anatase (101). Indeed,
our calculations for the oxygen vacancies ($V_{\rm O}$) on the two 
surfaces show that it is much harder to create $V_{\rm O}$ on anatase
compared to rutile. In particular, the calculated $V_{\rm O}$
formation energy is 3.77 eV $+\Delta\mu_{\rm O}$
for rutile and 4.94 eV $+\Delta\mu_{\rm O}$ for anatase,
where $\Delta\mu_{\rm O}=\mu_{\rm O}-\frac{1}{2}\mu_{{\rm O}_2}$
is the deviation of the oxygen chemical potential from its molecular
value. These results are in good agreement with previous studies
and have been attributed by Diebold {\em et al.} to the lower coordination 
of the Ti atoms on anatase compared to rutile 
upon creation of $V_{\rm O}$.\cite{vo_form}
Such a large difference of 1.17 eV in $V_{\rm O}$ formation energies indicates
that for anatase and rutile samples prepared under similar conditions, 
surface oxygen vacancies will be significantly more abundant on rutile
(110) compared to anatase (101), which is in agreement with experimental 
results.\cite{Hebenstreit}
This result combined with the ordering of the calculated
BEs ($E_{R,{\rm Ana}}>E_{R,{\rm Rut}}>E_{S,{\rm Ana}}>E_{S,{\rm Rut}}$)
indicate that the 
concentration of oxygen vacancies must be the key controlling mechanism
for the observed preferential growth of Pt on rutile. That is, {\em if}
$V_{\rm O}$ {\rm had} similar formation energies on both rutile and anatase, 
one {\em would} expect to observe more Pt on anatase than on rutile (based on
$E_{R,{\rm Ana}}>E_{R,{\rm Rut}}$). However, since $V_{\rm O}$ is much less likely
to form on anatase under similar preparation conditions (determined by 
$\mu_{\rm O}$), the $V_{\rm O}$'s on rutile become the next preferred nucleation sites
for Pt particles as the BE of Pt on reduced rutile (3.59 eV)
is considerably larger than that on stoichiometric anatase (2.87 eV).

It is important to notice that all the analyses so far have been at a 
single atom level. Our experimental results, on the other hand, indicate that 
the average size of visible Pt particles is $\sim$ 1 nm, which will be composed of 
a few Pt atoms. We therefore took our investigation one step further 
by examining the binding of Pt$_2$ dimers to the reduced rutile and anatase
surfaces. The lowest energy atomic configurations of Pt$_2$ bound to 
$V_{\rm O}$ on rutile (110) and anatase (101) are shown in Figs. 2(e) and 2(f), 
respectively. On rutile, Pt$_2$ configuration seems very much like a 
combination of the structures of single Pt atoms on reduced and stoichiometric
surfaces, as one Pt is still at the substitutional site and the other at the 
hollow position with a Pt-Pt distance of 2.52 \AA. On anatase, a similar
configuration occurs with one Pt at the substitutional site and the other
in hollow position with a Pt-Pt bond at 2.50 \AA. The calculated BEs
per atom (Table I), 3.69 eV for rutile and 4.71 eV for anatase, 
reveal two interesting results. First, due to the increase in the BE per atom
from 3.59 eV to 3.69 eV, we conclude that a single oxygen vacancy on rutile (110)
can bind two Pt atoms, in contrast to the instability of Au$_2$ on the same
surface with respect to a single Au atom bound at $V_{\rm O}$ site and
a diffusing Au atom.\cite{Wahlstrom} Second, although the BE per atom on anatase
is still larger than that on rutile, BE/atom for Pt$_2$ on anatase 
has decreased by 0.34 eV compared to 
a single Pt atom, showing the instability of a Pt$_2$ cluster 
with respect to 2 separate
Pt atoms anchored at far away oxygen vacancies. Although we do not necessarily
expect that this trend will evolve monotonically as the number of Pt atoms
in the cluster increases (it does not, for Au clusters\cite{Wahlstrom}), the present 
results suggest that rutile (110) surface could trap larger Pt clusters
at oxygen vacancy sites compared to anatase. 
Indeed, although it is for the anatase (001) surface, 
it has already been reported that upon heating in
vacuum at between 300 and 670K, coalescence of Pt islands was
observed on rutile (110), but not on anatase
(001)-$(1\times 4)$.\cite{Gan}

In summary, we have performed a comparative first principles study of
Pt and Pt$_2$ on reduced and stoichiometric rutile (110) and anatase (101)
surfaces in order to address the microscopic origin of our recent experimental
finding of preferential growth of Pt on rutile TiO$_2$ rather than anatase.
Our results indicate that (i) Pt and Pt$_2$ bind more 
strongly to anatase surfaces compared to rutile, but oxygen vacancies as
the preferred nucleation sites of Pt clusters will be significantly more
abundant on rutile, and (ii) $V_{\rm O}$ on rutile can bind a Pt$_2$ cluster, 
while on anatase, Pt$_2$ is found to be unstable with respect to two 
isolated Pt atoms anchored at $V_{\rm O}$ sites. Therefore, we conclude
that the selective growth of Pt on rutile must
be controlled by the relative number of available nucleation
sites ($V_{\rm O}$) and the seemingly stronger 
tendency of $V_{\rm O}$ on rutile to trap more 
Pt particles compared to anatase. 
Incorporation of larger Pt clusters and their diffusion are
necessary to further improve our understanding of the possible phenomena
behind the SMSI and the dense clustering of Pt on rutile.
This work was supported by the ACS Petroleum Research Fund under
grant \#s 40028-AC5M and 37552-AC5, and by NCSA under grant \#
DMR030053.


\begin{thebibliography}{99}

\bibitem{Diebold} For an excellent review, see
U. Diebold, Surf. Sci. Rep. {\bf 48}, 53 (2003) and references therein.

\bibitem{Tauster} S. J. Tauster, S. C. Fung, and R. L. Garten, J. Am. Chem.
Soc. {\bf 100}, 170 (1978).

\bibitem{Li et al.} Y. Li, Y. Fan, H. Yang, B. Xu, L. Feng, M.
Yang, Y. Chen, Chem. Phys. Lett. {\bf 372}, 160 (2003).

\bibitem{Fu-gra-had-ohno} A. Fujishima and K. Honda, Nature (London)
{\bf 238}, 37 (1972); M. Gr\"{a}tzel, Comments Inorg. Chem.
{\bf 12}, 93 (1991); K. I. Hadjiivanov and D. K. Klissurski,
Chem. Soc. Rev. {\bf 25}, 61 (1996); T. Ohno, K. Sarukawa, and M. Matsumura,
J. Chem. Phys. B. {\bf 105}, 2417 (2001).

\bibitem{hurum} D. C. Hurum, K. A. Gray, T. Rajh, and M. C. Thurnauer, 
J. Chem. Phys. B {\bf 109}, 977 (2005).

\bibitem{Iddir1} H. Iddir, M. M. Disko, S. \"{O}\u{g}\"{u}t, and N. D. Browning,
Micron {\bf 36}, 233 (2005).

\bibitem{Launay} M. Launay, F. Boucher, and P. Moreau,
Phys. Rev. B {\bf 69}, 035101 (2004).

\bibitem{Rut-akita} P. Ruterana, P.-A. Buffat, K. R. Thampi, M. Graetzel, Mat.
Mat. Res. Soc. Symp. Proc. Vol. {\bf 139}, 327 (1989); 
T. Akita, M. Okumura, K. Tanaka, S. Tsubota and M.
Haruta, J. Elect. Microscopy {\bf 52}, 119 (2003); T. Akita, P. Lu,
S. Ichikawa, K. Tanaka, and M. Haruta, Surf. Interface Anal. {\bf 31},
73 (2001).

\bibitem{vasp} G. Kresse and J. Hafner, Phys. Rev. B {\bf 47}, R558 (1993);
G. Kresse and J. Furthm\"{u}ller, {\em ibid.} {\bf 54}, 11169 (1996).

\bibitem{Rama-Laz} M. Ramamoorthy, D. Vanderbilt, and R. D. King-Smith,
Phys. Rev. B {\bf 49}, 16721 (1994); M. Lazzeri, A. Vittadini and, A. Selloni, 
{\em ibid.} {\bf 63}, 155409 (2001);  {\em ibid.} {\bf 65}, 119901 (2002).

\bibitem{Iddir2} H. Iddir, S. \"{O}\u{g}\"{u}t, N. D. Browning, and M. M. Disko
Phys. Rev. B {\bf 72}, 081407(R) (2005). The Pt binding energies reported
in this reference should actually be 0.4 eV lower, which does not affect 
the main conclusions, and has been recently submitted as an Erratum.

\bibitem{note} On both stoichimetric rutile and anatase surfaces,
the Pt bonding to Ti atoms is covalent as indicated by the
appreciable charge build up between Pt and Ti, and the Pt    
bonding to oxygen is ionic with a considerable charge polarization along
the Pt-O bond direction (On anatase, there is, however, a small amount
of charge build up between Pt and O which indicates partial covalency).
On the reduced surfaces, Pt bonding to the nearest two Ti atoms is 
strongly covalent with a significant charge build up between Pt and Ti atoms.

\bibitem{karin} X. Huang, K. M. Rabe, and G. J. Ackland, Phys. Rev. B {\bf 67}, 
024101 (2003).

\bibitem{vo_form} A. Vittadini and A. Selloni,
J. Chem. Phys. {\bf 117}, 353 (2002); X. Wu, A. Selloni, and S. K. Nayak,
{\em ibid.} {\bf 120}, 4512 (2004); M. D. Rasmussen, L. M. Molina,
and B. Hammer {\em ibid.} 988 (2004); A. Vijay, G. Mills, and H. Metiu,
{\em ibid.} {\bf 118}, 6536 (2003); T. Bredow and G. Pacchioni, Chem. Phys.
Lett. {\bf 355}, 417 (2002); U. Diebold, N. Ruzycki, G. S. Herman, and A.
Selloni, Catal. Today, {\bf 85}, 93 (2003).

\bibitem{Hebenstreit} W. Hebenstreit, N. Ruzycki, G. S. Herman, Y. Gao, 
and U. Diebold, Phys. Rev. B {\bf 62}, R16334 (2000).

\bibitem{Wahlstrom} E. Wahlstr\"{o}m, N. Lopez, R. Schaub, P. Thostrup, A.
R\o nnau, C. Africh, E. L\ae gsgaard, J. K. N\o rskov, and F.
Besenbacher, Phys. Rev. Lett. {\bf 90}, 026101 (2003).

\bibitem{Gan}
S. Gan, A. El-azab, Y. Liang, Surf. Sci. {\bf 479}, L369 (2001).

\end{thebibliography}
\end{document}